# Implementation of Big Data Analytics for Diabetes Management: Needs Assessment in the Rwanda Healthcare System

Silas Majyambere
Department of Computer and Systems Sciences
Stockholm University
Stockholm, Sweden
E-mail: majyambere@dsv.su.se

Workneh Yilma Ayele
Department of Computer and Systems Sciences
Stockholm University
Stockholm, Sweden
E-mail: workneh@dsv.su.se

Tony Lindgren
Department of Computer and Systems Sciences
Stockholm University
Stockholm, Sweden
E-mail: tony@dsv.su.se

Celestin Twizere
Center for Biomedical Engineering and E-Health (CEBE)
University of Rwanda
Kigali, Rwanda
E-mail: celestintwizere@gmail.com

***Abstract*—Diabetes is a chronic metabolic disease that can lead to serious health problems if not diagnosed and managed early. Big Data Analytics (BDA) and machine learning offer practical tools for analyzing large health datasets and supporting early detection and better treatment decisions. However, their use in routine clinical practice is still limited. This study examines the readiness of Rwanda's healthcare system to adopt big data analytics for diabetes management. As the country continues to expand its use of electronic medical records and health information systems, new opportunities arise for improving prediction, monitoring, and clinical decision-making. A five-day workshop involving 25 key stakeholders, including clinicians, data managers, policymakers, medical researchers, nutritionists, and technology providers, was conducted to assess preparedness and identify existing gaps. The findings highlight both the potential and the main challenges of BDA implementation. Based on these results, the paper proposes a practical BDA framework to support diabetes management strategies using explainable machine learning models**.

***Keywords-Big Data; Big Data Analytics; Diabetes Management; Explainable Machine Learning; Rwanda Healthcare System*.**

## I. INTRODUCTION

The prevalence of diabetes cases continues to rise globally [1], prompting a call for action to mitigate healthcare expenditure on diabetes management. Diabetes is a global health burden characterized by elevated blood glucose levels [1]. It is a life-threatening disease in the sense that the uncontrolled blood glucose leads to devastating complications that come with damage to different body organs, followed by chronic diseases such as macrovascular and microvascular diseases. The most frequent diabetic macrovascular diseases are heart diseases, coronary artery disease, and stroke, and diabetic microvascular diseases, which are under study in this paper, are diabetic retinopathy, neuropathy, diabetic foot ulcers, and nephropathy [2]. Research studies have demonstrated that uncontrolled diabetes has a positive association with neurodegenerative diseases, including Dementia and Alzheimer's disease [3][4]. There are three major types of diabetes [5]. Type 1 diabetes is characterized by the pancreas's inability to produce the required insulin effectively. Type 2 diabetes is characterized by the body's ineffective use of insulin. There is also gestational diabetes, which consists of high blood glucose during pregnancy. Insulin is essential for the processing of glucose and its storage in cells as a source of energy.

Diabetes has been extensively studied to identify its risk factors and delay complications [6]. More recently, health informatics and data science researchers have focused on predicting diabetes using electronic health record data [7]. Early detection benefits both patients and care providers by enabling timely treatment, preventing or delaying end-stage complications, and reducing healthcare costs. However, machine learning models trained on small datasets often perform well during development but poorly in real-world healthcare settings and lack clinical interpretability [7][8]. Big data, defined as the application of advanced data mining techniques to discover patterns from large datasets [9], has the potential to advance value-based or precision medicine by supporting evidence-based decision-making [10]. Although various machine learning algorithms achieve high predictive accuracy, limited attention has been given to healthcare providers' readiness to adopt these models for effective big data implementation [11]. A gap, therefore, remains between theory and practice in the successful implementation of big data analytics for diabetes management. This study evaluates healthcare providers' preparedness in Rwanda to adopt big data analytics for diabetes management.

### *A. Healthcare System in Rwanda*

In Rwanda, the public healthcare system dominates the health services and works in a decentralized manner. Health service delivery begins with community health workers [12].

These Community Health Workers (CHWs) are trained volunteers without formal medical education who receive targeted training and continuous support from the Ministry of Health (MoH) to help identify and manage conditions such as malaria, diabetes, and hypertension. Equipped with smartphones, CHWs collect patient information using *mUzima* [13], a mobile health (mHealth) application integrated with OpenMRS, the electronic health record system implemented across most public health centers. The structure of public healthcare is presented in Figure 1.

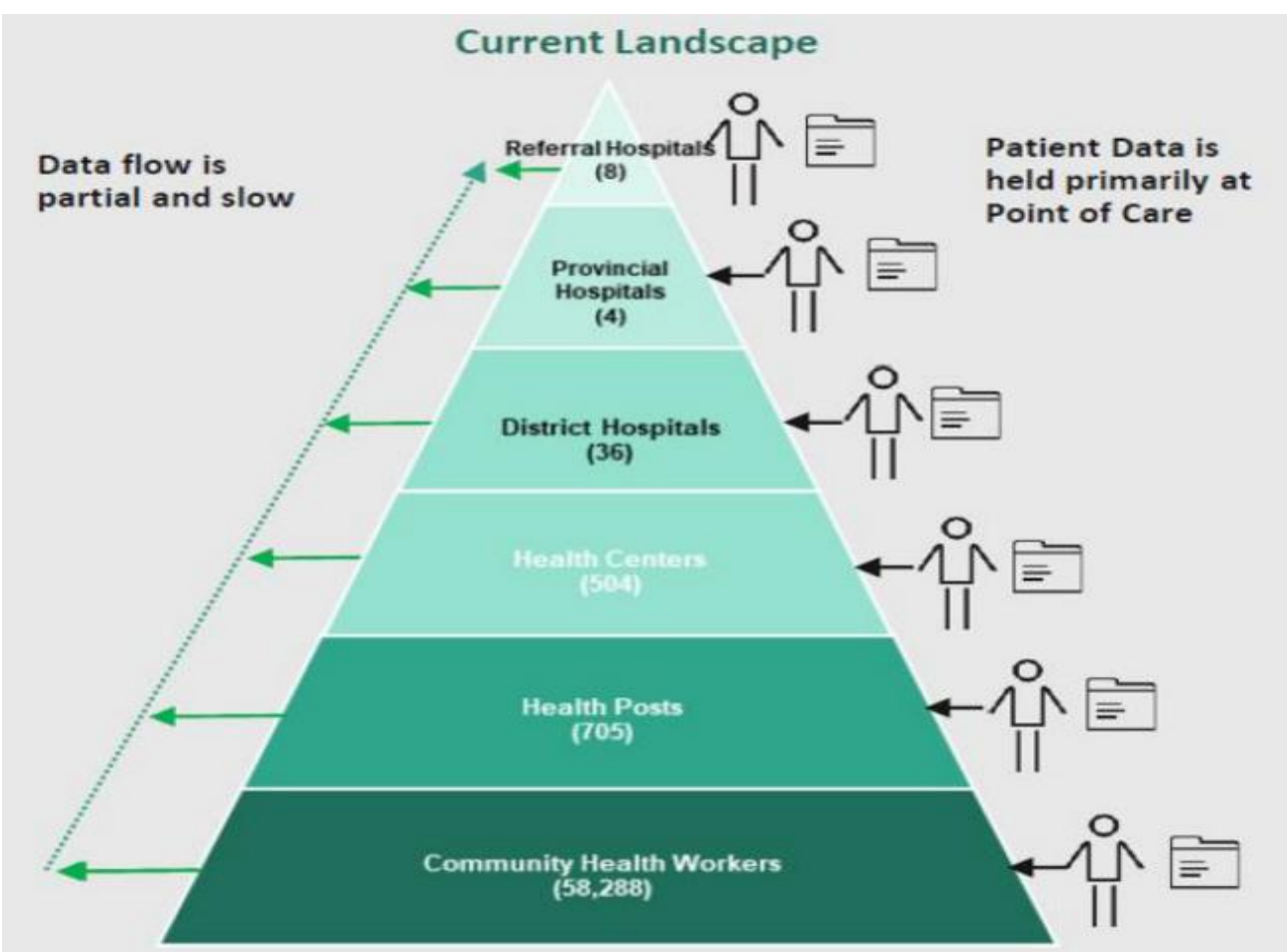

Figure 1. Structure of Public Healthcare in Rwanda.

The flow of the clinical decision-making process is outlined in Figure 2. Health posts, located in close proximity to communities, provide basic healthcare services and serve as the primary point of contact for many patients. Health centers, the most frequently utilized healthcare facilities, report to district hospitals, which provide general medical services. Patients requiring specialized care are referred to referral hospitals. Referral hospitals interact with the Ministry of Health (MoH) and its affiliated institutions, including the Rwanda Biomedical Center (RBC).

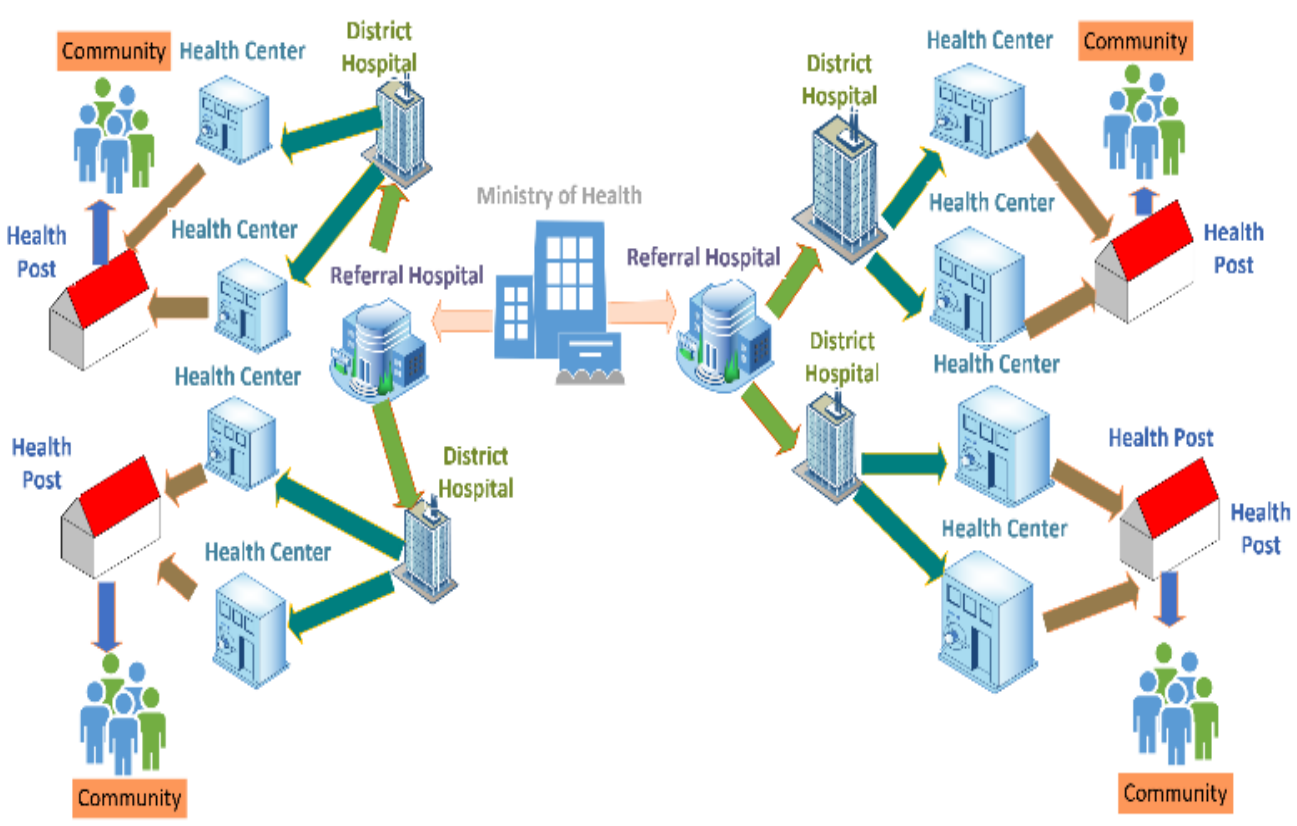

Figure 2. Flow of clinical decision-making.

Diabetes care services are delivered across all levels of the healthcare system, including diagnosis, scheduled visits, medication, and follow-up management, primarily conducted at health centers and district hospitals.

## B. *Rwanda's Current Advancements in Digital Health*

Rwanda has undergone a significant transformation in its healthcare sector [14][15], driven by substantial investments in infrastructure and the adoption of Information and Communication Technologies (ICTs) to enhance digital health and improve service delivery. Key reforms have included expanding the healthcare workforce, decentralizing healthcare services, implementing universal health coverage through the Community-Based Health Insurance (CBHI) program, which allows citizens to access care at an affordable cost, and introducing a unified patient data as demonstrated in Figure 3.

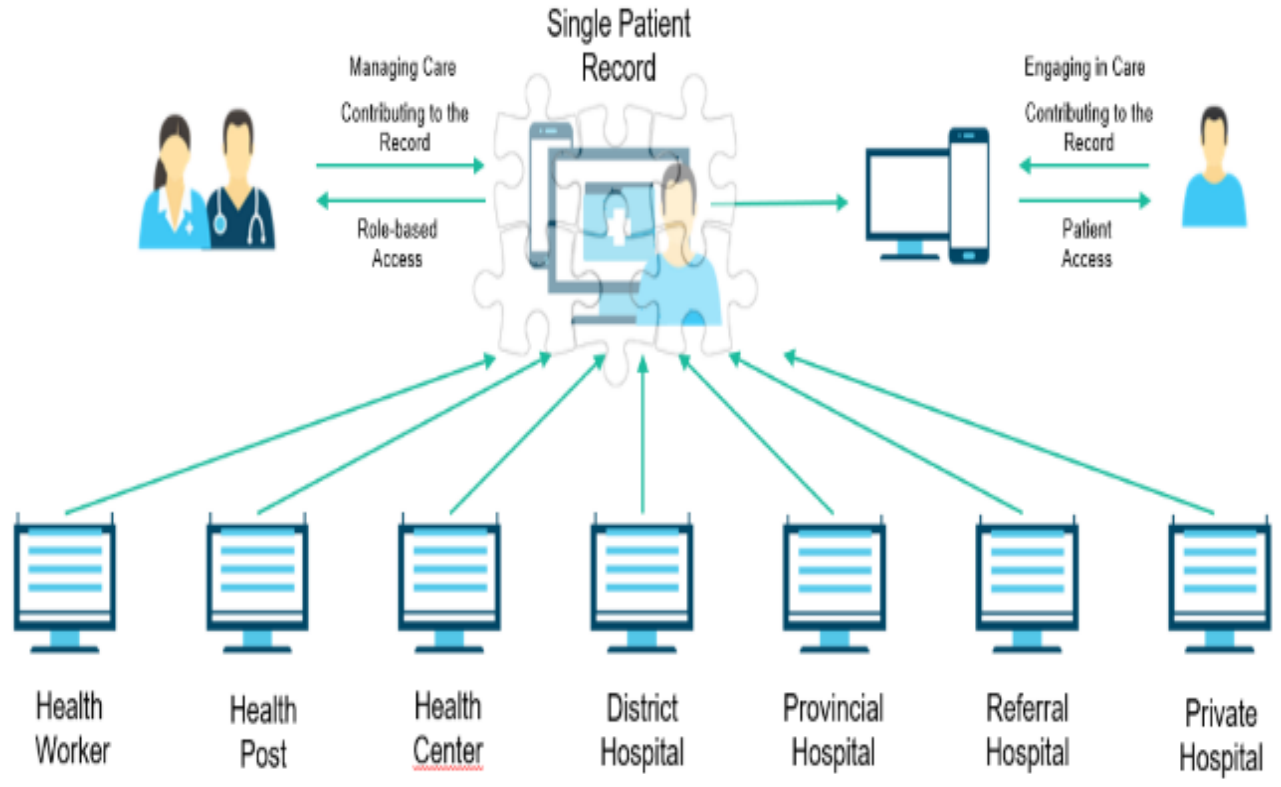

Figure 3. Unified patient records.

The single patient record is achieved through data integration and interoperability of healthcare systems. This will unlock the potential of big data analytics to improve healthcare service delivery and control complex diseases, such as diabetes, through patient profiling and tracking disease progression. Figure 4 illustrates the interconnectivity of healthcare systems utilizing the Rwanda Health Cloud, demonstrating Rwanda's commitment to leveraging the latest technology in the digital transformation of the healthcare sector.

Open Medical Record System (OpenMRS) [16] and OpenClinic [17] are the primary Electronic Medical Record (EMR) systems used across public and private health facilities in Rwanda. A major milestone in the country's digital health advancement is the plan to use the Health Information Exchange (HIE), which supports interoperability and enables a "one patient, one record" approach across all levels of care.

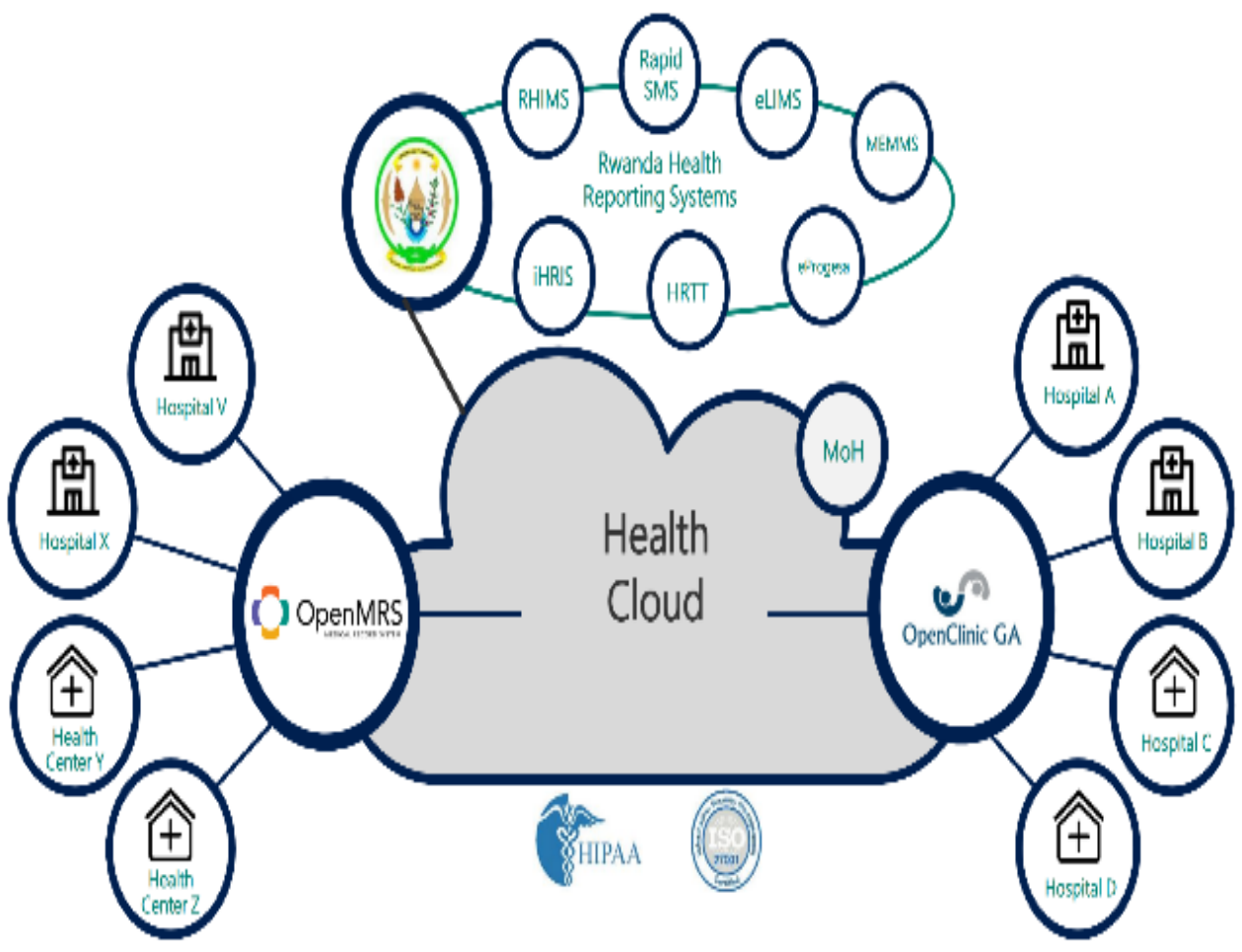

Figure 4. Rwanda Health Cloud.

*C. Diabetes Management in Rwanda*

In Rwanda, diabetes patients make regular visits every three months. During each visit, physical examinations and modifiable risk factors are assessed, and the data are recorded in the electronic health records keeping systems. The purpose of the regular visits is to check the quality of health of diabetic patients, and it is a way of tracking the progression of the disease and assessing the side effects of medication in order to prevent or delay the complications associated with diabetes progression, such as heart disease and diseases that affect the nerves, eyes, and kidneys. Healthcare providers use a wide range of electronic devices to access and record data from diabetic patients. Currently, there is a lack of flexible methods for analyzing disease dynamics using data available in health information systems. This research contributes to the assessment of current infrastructure and skills to adopt big data analytics methods of knowledge discovery from large data of diabetic patients in Rwanda and we propose a framework based on machine learning techniques to track diabetes progression with meaningful data visualization and accurate predictions, the new framework works as an intelligent decision support system which differ from the existing systems by adding a layer of analytics and visualization.

*D. BDA in Diabetes Management Literature*

As diabetes data continues to grow in scale and diversity, it opens new avenues for research to examine the role of big data analytics in diabetes management. Harnessing these rich datasets has the potential to significantly enhance our understanding of disease dynamics and improve the quality of care for the diabetic population. A study examined the benefits of combining big data analytics with machine learning for medical decision-making [9]. Using MLlib, the machine learning library in Apache Spark, the authors trained and evaluated predictive models on a large dataset from the Behavioral Risk Factor Surveillance System (BRFSS), comprising 441,455 samples and 330 features. Logistic Regression achieved the highest performance, reaching an accuracy of 75%. The study explores the benefits of big data and artificial intelligence [18].

The Italian Association of Medical Diabetologists (AMD) demonstrates a positive commitment to adopting these technologies to enhance diabetes care through comprehensive therapeutic interventions supported by evidence from data-intensive analytics tools. This will help mitigate the shortages of experienced diabetologists and the increasing number of diabetes patients. A scoping review conducted by [19] highlights the key characteristics of big data relevant to diabetes management, framing them within the well-established 5V model: volume, variety, velocity, veracity, and value. These dimensions highlight the complexities involved in extracting meaningful insights from the large, heterogeneous, rapidly generated, and often distributed data from diabetic patients. Such insights are critical for supporting descriptive, diagnostic, predictive, and prescriptive analytics in diabetes care. The authors argue that integrating big data and artificial intelligence holds considerable potential to improve diabetes management, including the ability to better understand factors contributing to poor glycemic control in individual patients.

Researchers in [20] have developed a big data analytics model to optimize medication for diabetic patients with comorbidities or comedications. This model was validated on a dataset of 19,223 samples. The model uses the concept of Markov blankets in Bayesian networks to simplify the computational challenges of optimization in complex datasets, such as the diabetes dataset, which combines clinical, demographic, and medication features in the presence of comorbidities. The model achieved a precision of 88.75% and an Area Under the Curve (AUC) of 71.15%. The article [21] synthesizes insights from a panel of experts in diabetes management, emphasizing the importance of integrating patient-level diabetes data to enable effective information extraction. The authors argue that such integrated diabetes data can provide timely, actionable guidance to policymakers, healthcare administrators, clinicians, and the public, particularly for short-term decision-making and long-term planning to control the disease. They further believe that enhanced integration and management of big data can create new opportunities to advance diabetes care and research.

The rest of the paper is organized as follows: Section 2 explains the qualitative data collection and analysis; Section 3 presents the focus group results; Section 4 explores healthcare stakeholders' views on using big data for diabetes management in Rwanda; and Section 5 concludes with future research directions.

## II. METHODOLOGY

The aim of this research is to evaluate Rwanda's readiness to adopt big data analytics for nationwide diabetes management. Specifically, the study investigates the foundational requirements for implementation, including digital infrastructure, workforce capacity, and data availability, and examines the potential contribution of explainable machine learning analytics tools. To gain insight into the current understanding and use of big data in disease management, we conducted a purposive group discussion focused on diabetes as a case example. Diabetes in Rwanda generates substantial clinical data and remains one of the most expensive chronic conditions for both patients and healthcare providers. As such, it represents a high-impact domain in which advanced data analytics could significantly enhance disease monitoring, management, and overall care delivery. Focus Group (FG) discussions are widely used qualitative data collection methods in health research [22][23]. Purposive sampling was employed to recruit participants, as selecting knowledgeable individuals is critical to generating meaningful discussion outcomes. For this study, the FG was conducted as a five-day workshop. This approach was chosen for its interactive nature, which facilitates the efficient collection of participants' perceptions, attitudes, and beliefs regarding the research topic. The discussion was moderated by two PhD students conducting research on Diabetes and Hypertension Management using Technology. The workshop was conducted in English.

### A. Recruitment of Participants

Participant selection was guided by anticipated expertise in diabetes and hypertension management, the use of technological applications in chronic disease care, nutrition, health policy, and ICT policy within the healthcare sector. Given the study's relevance and importance, the workshop was conducted in a hybrid format, combining both in-person and online participation. The number of participants was determined by the available budget, which needed to cover transportation and lodging costs for in-person attendees. An Email explaining the purpose of the workshop was sent to 30 participants. Among 30 Emails sent, 25 participants confirmed to attend the workshop in face-to-face mode (18) and online mode (7). For those who decided to participate in person, a formal invitation was sent to their respective institution. The workshop was conducted in Musanze District from June 24 to 28, 2024, at Virunga Hotel. All expenses related to the workshop were covered by the University of Rwanda (UR), the Center of Excellence in Biomedical Engineering, and the E-Health in support of the UR-Sweden Program for capacity building and research partnership.

### B. The Purpose of Group Discussion

To assess Rwanda's readiness to implement big data analytics in diabetes management, we convened a workshop-style discussion with diabetes specialists, hospital data managers, healthcare technology providers, senior medical researchers, health policymakers, nutritionists, and ICT policymakers in Rwanda. The discussion covered knowledge sharing on diabetes, its associated complications, the impact of hypertension on diabetes patients, how to control its progression, and how emerging technologies (such as big data analytics) can leverage diabetes data to improve its management. The benefits and challenges of implementing big data for diabetes management in Rwanda were also discussed.

### C. Characteristics of participants

Participants were organized into eight professional subgroups based on their areas of expertise: Senior Medical Researchers (SMR, n=4), Medical Practitioners (MPR, n=5), Nutritionists (NUT, n=3), Medical Technology Providers (MTP, n=3), Data Managers (DMA, n=3), Health Policy Makers (HPM, n=3), ICT Policy Makers (IPM, n=2), as well as researchers specializing in Diabetes (DIA, n=1) and Hypertension (HYP, n=1). Detailed participant characteristics are presented in Table I.

TABLE I. CHARACTERISTICS OF PARTICIPANTS

| Subgroup | Participants | Gender | Age group | | Coding |
|---|---|---|---|---|---|
| **SMR** | 4 | F:1, M:3 | 50-60: 3, >60: 1 | | SMR1-4 |
| **MPR** | 5 | F: 1, M: 4 | 20-29: 1, 30-39: 2, 40-49: 2 | | MPR1-5 |
| **NUT** | 3 | F: 1, M: 2 | 20-29: 1, 30-39: 2 | | NUT1-3 |
| **MTP** | 3 | F: 1, M: 3 | 30-39: 1, 40-49: 2 | | MTP1-3 |
| **DMA** | 3 | F: 0, M: 3 | 20-29: 1, 30-39: 2 | | DMA1-3 |
| **HPM** | 3 | F: 0, M: 3 | 40-49: 2, 50-59: 1 | | HPM1-3 |
| **IPM** | 2 | F: 0, M: 2 | 40-49: 2 | | IPM1-2 |
| **DHR** | 2 | F: 1, M: 1 | 40-49: 2 | | DIA, HYP |

### D. Group Discussion Moderation

The focus group discussion was facilitated by two moderators with expertise in diabetes (DIA) and hypertension (HYP). During the first two days of the workshop, each subgroup was invited to present its routine professional activities and describe how its work relates to the management of diabetes and hypertension. Guided by the moderators, a set of core discussion topics was identified for the remaining three days. These topics included: diabetes screening; diabetes-related complications; medical data management and governance; existing resources and requirements for implementing big data analytics in diabetes management; the role of big data analytics in diabetes and hypertension care; the contribution of nutrition to managing these conditions; and the potential of explainable machine learning to enhance clinical adoption of big data analytics techniques.

### E. Data analysis

The thematic analysis approach was employed to analyze the data collected from participants in the group discussions during a five-day workshop organized by the Center of Excellence in Biomedical Engineering and E-Health (CEB).

## III. RESULTS

During the assessment of current practices for diagnosing diabetes, we summarized the outcomes of diabetes diagnosis across different stages (Figure 5). As illustrated in Figure 5, the complexity of diabetes management increases significantly once the disease is confirmed, progressing toward various complications influenced by multiple risk factors, including lifestyle, aging, and environmental determinants.

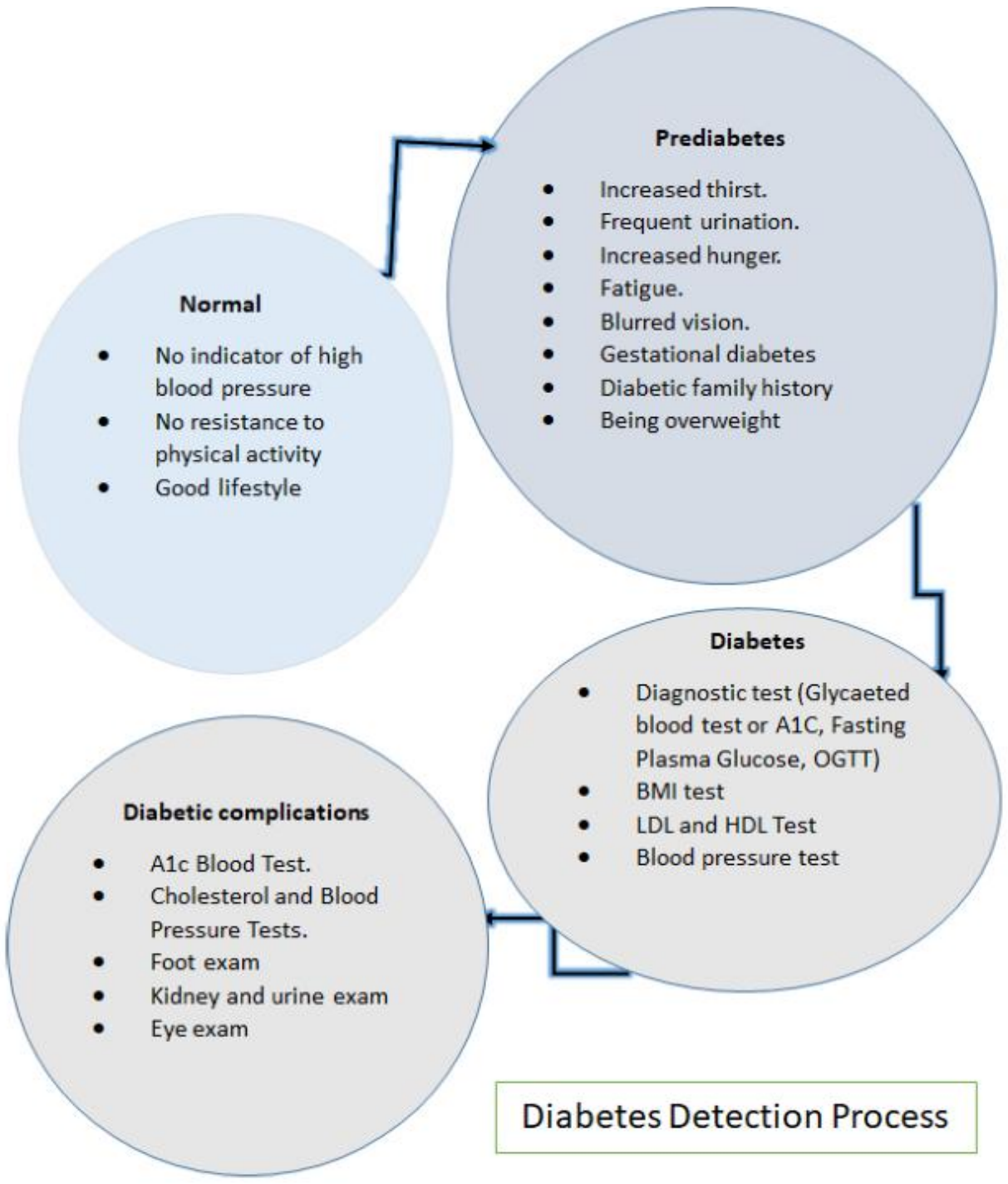

Figure 5. Process of Diabetes Diagnosis.

In Rwanda, disease progression is monitored through routine follow-up visits; however, diabetes specialists reported that effectively tracking patients at risk of developing complications remains challenging without the support of advanced big data analytics tools. Participants also highlighted limited coordination and data sharing among the clinical departments involved in diabetes care. This gap was captured by one medical practitioner, who noted: *"It is not clear how patients reach proliferative retinopathy, leading to blindness, while this patient regularly visits the internal medicine department"* (MPR3). All participants acknowledged the potential benefits of integrating big data into current diabetes management practices. "*It is high time to use big data analytics to investigate the effect of a healthy diet by consuming food with a low glycemic index or a diabetes-friendly meal on diabetes management and medication*", said NUT2.

One participant, diagnosed with diabetes in 2010, shared challenges related to limited knowledge about disease progression and management. "*Diabetes is a silent killer disease; it gradually damages our organs and weakens the body over time. With recent technologies like ChatGPT, we can better understand it and how to live better with it, but sometimes the content generated by ChatGPT can be misleading. I believe the ideas gathered here can be turned into a simple, effective tool to improve our health and strengthen interaction between patients and healthcare providers. As a patient, I'm ready to use locally developed big data tools to help address the challenges of diabetes*," said DMA2.

Based on discussions about data availability, barriers to data access and sharing, and the types of medical analytics that big data can support, an overall perspective on how big data could be implemented in the Rwandan context is presented in Figure 6.

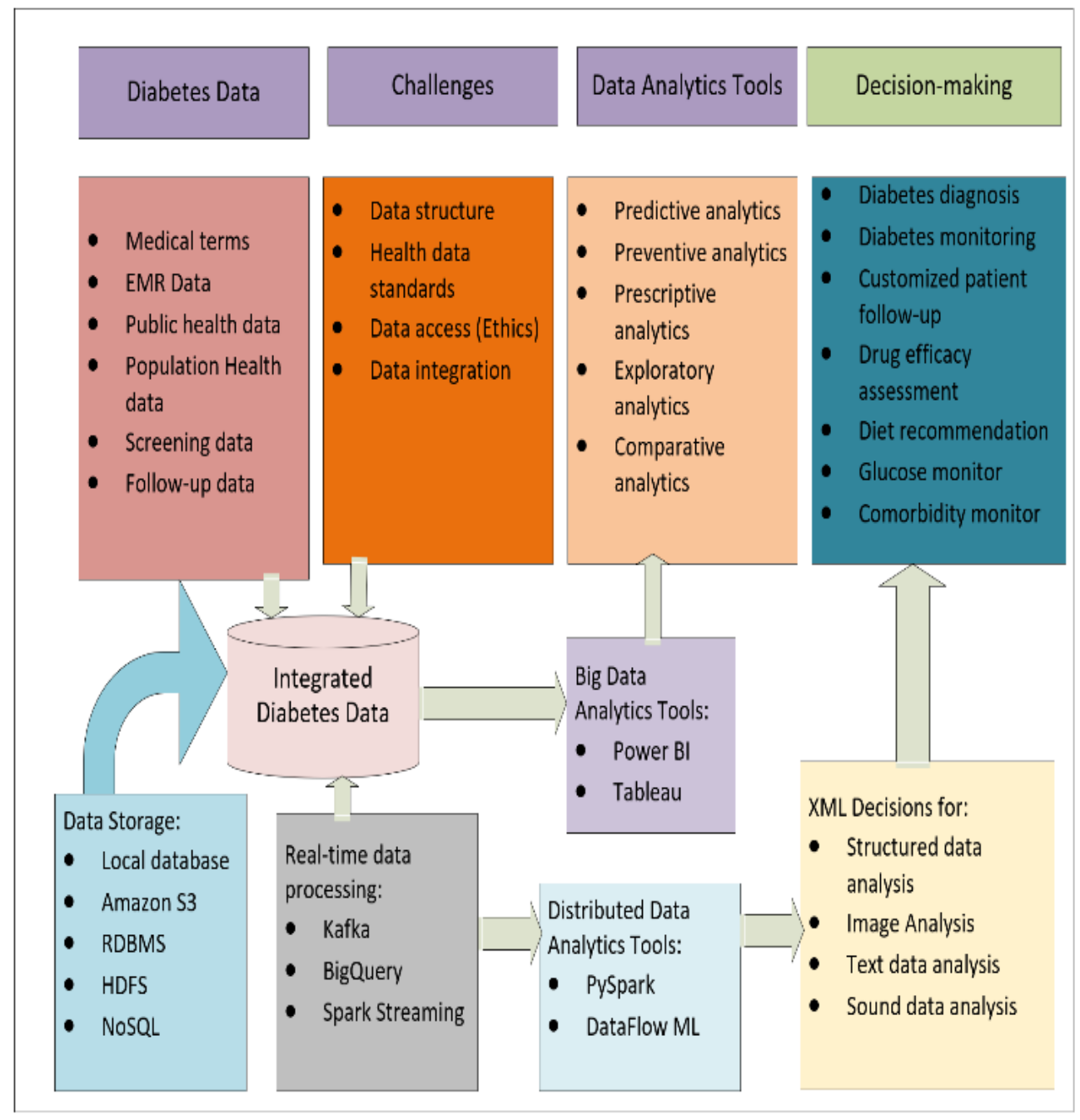

Figure 6. Architecture of Big data for Diabetes management.

Integrated diabetes data form the foundation for applying big data analytics. Once consolidated, the large volume of diabetes-related information can be extracted, processed, and used to develop Explainable Machine Learning (XML) models, which depend on high-quality input data. These models enable various forms of analytics and serve as decision-support tools to help clinicians make more informed clinical judgments, thereby improving overall diabetes care.

### A. *Major themes identified*

The result of thematic analysis indicates that there are five main themes from the analysis of the topics discussed during workshop-based group discussion: (1) Big data analytics promises, (2) Big data analytics challenges, (3) Data governance, (4) Big data analytics quality standards, and (5) Clinical data intelligence. Through deeper analysis of themes and a general overview of the big data ecosystem, we develop a framework for managing diabetes that can be adapted to other chronic diseases by using the same XML models, but trained on datasets specific to each chronic disease. The framework is illustrated in Figure 7.

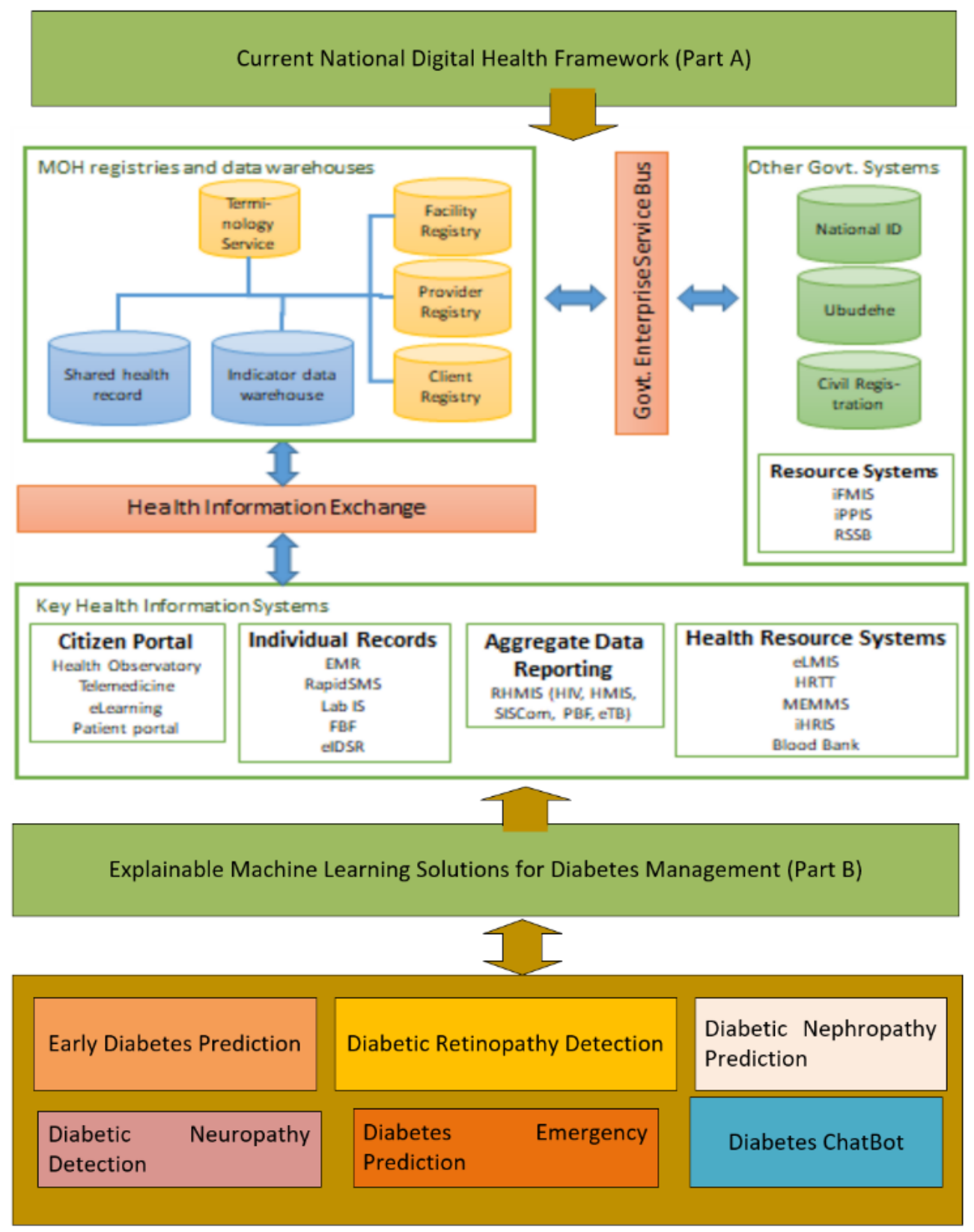

Figure 7. Proposed Big Diabetes Data analytics integration in the Rwanda Healthcare System.

The proposed framework extends the national digital health infrastructure (Part A) [14] with an additional layer of Explainable Machine Learning (XML) solutions for diabetes management (Part B). This layer leverages aggregated patient data to develop a suite of machine learning models that support diabetes management in several key areas. First, the XML models enable early prediction of diabetes by analyzing demographic information, symptoms, vital signs, and laboratory results obtained during previous diagnostic encounters. These models help identify individuals at increased risk of developing diabetes, thereby facilitating timely interventions. Following a confirmed diagnosis, the framework incorporates a Diabetes ChatBot powered by Large Language Models (LLMs) with Retrieval-Augmented Generation. By grounding responses in validated, context-specific data rather than relying solely on generative outputs, the system reduces the risk of hallucinations.

This ChatBot provides patients with guidance on managing blood glucose levels, understanding potential complications, and implementing preventive strategies through healthy dietary practices, regular physical activity, and appropriate glycemic targets. Additionally, deep learning models, particularly Convolutional Neural Networks (CNNs), are employed to detect diabetic retinopathy using retinal images acquired through Optical Coherence Tomography or smartphone-based fundus imaging. CNN models can detect diabetic retinopathy and grade its severity stages. Other specialized XML models address the detection of diabetic neuropathy and diabetic kidney disease using patient historical data. The use of explainable machine learning across these components ensures that predictive outputs are interpretable, thereby supporting clinician trust and informed decision-making. Overall, the proposed framework enhances clinicians' ability to understand the dynamics of diabetes progression and make evidence-based decisions, ultimately improving patient outcomes. It also empowers patients to adopt diabetes-friendly lifestyles and adhere to follow-up plans more cost-effectively.

### B. *Promises of Big Data for Diabetes Management*

During the focus group discussion, participants described their experiences using existing technologies to deliver patient care and expressed appreciation for the Government of Rwanda's ongoing ICT investments and the support provided by ICT departments, where data managers are based. They indicated strong openness to adopting new technologies to enhance clinical efficiency and reduce their workload. Figure 8 summarizes participants' expectations regarding how big data analytics could improve current diabetes care practices.

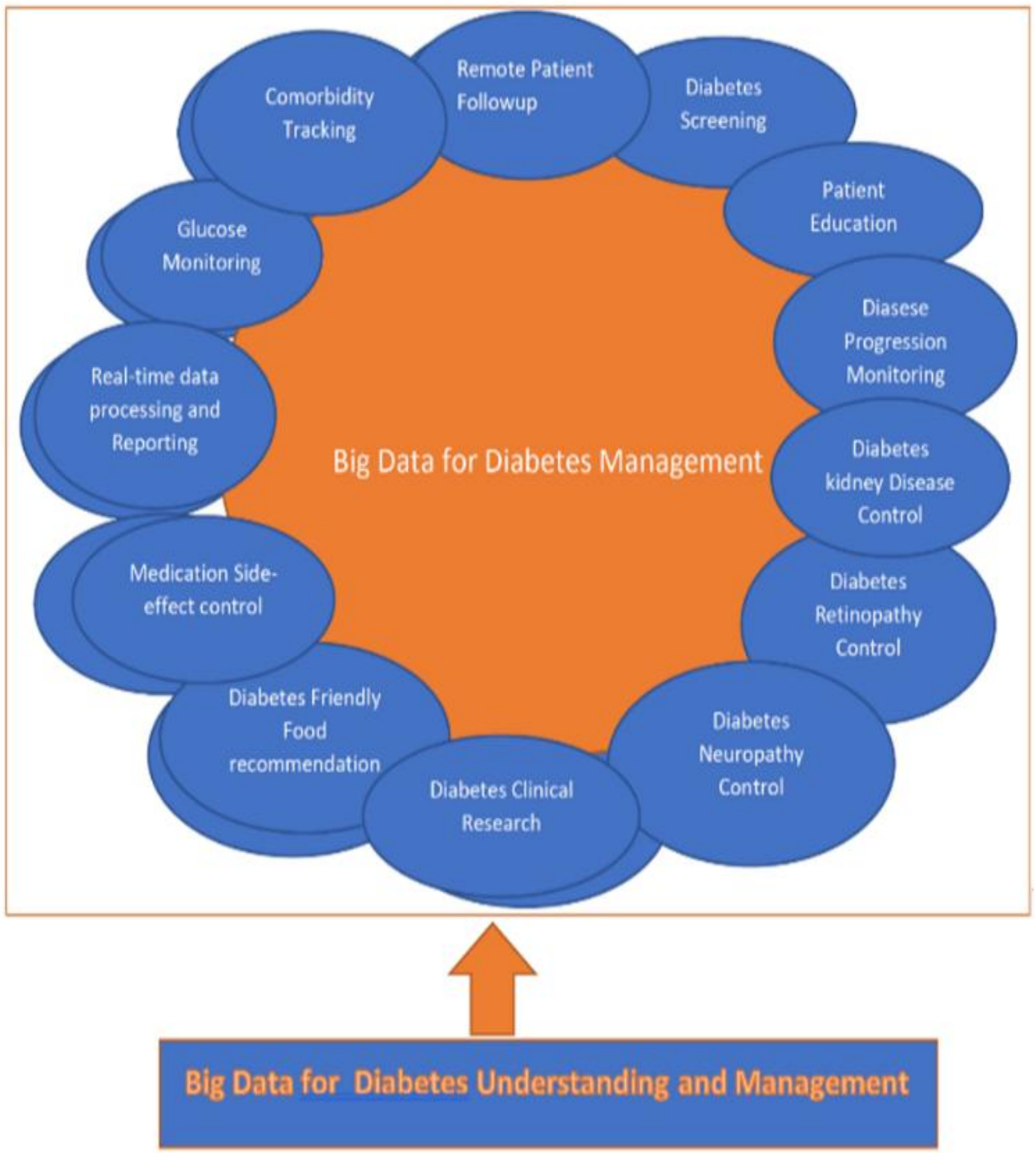

Figure 8. Big data analytics for Diabetes management.

### C. *Challenges Identified*

Healthcare systems in Rwanda have established good practices in recording patient data; however, the participants identified the following challenges that need to be addressed before big data analytics can have a positive impact on healthcare in Rwanda: (1) Patient data sharing. Data managers noted that sharing data between District hospitals and referral hospitals remains a significant challenge. Additionally, private hospitals utilize various health information systems and do not share data with public hospitals. (2) There is a need for harmonization of medical records to facilitate the implementation of big data analytics, which has the promising capability of analyzing massive healthcare data and coming up with insightful patterns in data. (3) There is a need to review the policy governing healthcare ethical requirements to mitigate the challenges of accessing and using patient data in Rwanda. (4) Big data analytics in healthcare is a dream technology for handling the complexity of chronic diseases, including diabetes; however, the gap in skilled people can hinder its successful implementation. (5) Instead of ready-made big data analytics systems, the consideration of input from domain experts in clinical settings, from the design to the deployment of such systems, can result in explainable and interpretive machine learning-based systems. The power of big data analytics in healthcare lies in its usage of data from different sources and formats, and using the four advanced analytics techniques for descriptive, predictive, diagnostic, and prescriptive methods using machine learning to produce insightful output that can generate evidence-based decisions.

## IV. DISCUSSION

We conducted the study on preparedness of healthcare providers to embrace the power of big data analytics to address the complexity of chronic diseases diagnosis and their treatment, the aim of this study is to assess the current ICT infrastructure, data management and the level of expertise required in healthcare to use the big data analytics to efficiently and accurately diagnose the diabetes and its complications using available data stored in the electronic health records systems.

All participants agree that the use of electronic medical records has improved the quality of healthcare service delivery, the policy makers have a bigger picture of transforming the healthcare to include the big data predictive analytics to promote big value of care and personalized medicine to address the chronic diseases and they are not far from the rest of world [24] on the promises of big data analytics to improve clinical outcomes, however the data managers highlight the lack of data analytical tools for mining the useful patterns from the data and more attention can be on how to build quality datasets and the medium to share the data.

Diabetes specialists need an easy way of screening for diabetes, diagnosing diabetes, and tracking diabetes progression to be able to prevent or delay its devastating complications. They record too much data on diabetes, but given the current information systems in place and the level of ICT knowledge they have, they are not efficiently benefiting from it. *"I believe that relying solely on clinical outputs from big data analytics is not a good idea, particularly when human expertise is not considered during the construction of these systems*," said MPR3.

### A. Challenges and Opportunities of Big Data analytics in healthcare

The results indicate that Rwanda is ready to use big data analytics to handle diabetes however the participants highlighted the challenges that need to be addressed such as harmonizing medical records, building high quality datasets, handling ethical issues related to access and use of patients' data, need of skilled people in data science, training of healthcare professionals on the use of big data analytics systems and interpreting the results generated by these systems. Some of these challenges, including data quality, data sources, tool selection for big data analytics, and the interpretation of predictive analytics results, were reported in [25]; however, the power of big data analytics in healthcare cannot be ignored. The participants suggested that big data should be well-designed to guide clinicians in tracking the trajectories of diabetes patients with clear visualization and recommendations after each medical visit, as well as to predict alarming situations within the diabetes population.

### B. Implementation Phases and Key Players

Participants in the group discussion workshop agreed that successful big data implementation should follow five phases. *Phase 1* focuses on data harmonization, which has already begun through the unification of patient data (see Figure 3). *Phase 2* addresses data storage and sharing. Key stakeholders, including the Ministry of Health (MoH), Ministry of ICT and Innovation (MinICT), Rwanda Information Society Authority (RISA), Rwanda Biomedical Center (RBC), and the Rwanda National Ethics Committee, have to collaborate to make large-scale diabetes datasets available for big data analytics research on the Rwanda Health Cloud. This phase also involves testing machine learning models designed to address specific diabetes-related challenges. These models should incorporate clinicians in the loop and provide explanations consistent with clinical knowledge. *Phase 3* involves piloting selected use cases in a referral hospital to evaluate feasibility and performance in a real clinical setting. *Phase 4* focuses on nationwide deployment, including training healthcare providers and guiding patients on the use of AI-based tools for diabetes management, ensuring informed consent, and integrating solutions into clinical workflows. *Phase 5* emphasizes continuous monitoring and evaluation of big data systems to detect potential bias, data drift, or changes in model performance. These oversight activities should be coordinated by the Rwanda Biomedical Center (RBC).

### C. Limitations

The study was conducted with a small sample, which limits generalizability to the broader population. The study did not consider the impact of ICT training on the daily duties of medical professionals since the electronic health records system is their working platform, and the level of skills required to understanding and using big data analytics systems was not assessed as such systems are not introduced yet in Rwandan healthcare; however, diabetes specialists have shown interest in participating in the design of a big data analytics framework by giving their ideas on how they think the system should assist them in making data-driven decisions on diabetes patient cases.

### D. Implications

Our study findings can serve as a starting point for implementing big data analytics in the healthcare work environment. To the best of our knowledge, no study has assessed the opinions of various healthcare stakeholders regarding the adoption of a medical big data analytics system. The identified challenges can be addressed by involving healthcare practitioners, data scientists, ICT policymakers, and diabetes researchers.

## V. Conclusion and Future Work

Big data analytics holds the potential to transform healthcare from expert-based decision-making to data-driven decision-making. Our study highlighted the opportunities and challenges of leveraging big data analytics in Rwanda. Our study suggests that Rwanda is poised to take the next step in implementing big data analytics once the aforementioned challenges are resolved. The participants expressed greater expectations for big data analytics to address the complexity of diabetes and its complications; however, the success of implementing the proposed framework depends on the collective effort of each healthcare stakeholder.

We suggest that diabetes care providers continue to document comprehensive patient information, as each data point contributes to the effectiveness of big data analytics. Given that diabetes management requires coordinated efforts, enhanced data sharing among clinical departments is essential to prevent progression to advanced disease stages, which increase healthcare costs and diminish patients' quality of life. Additionally, we suggest that the Ministry of Health (MoH) provide training for medical practitioners on the use of artificial intelligence in healthcare and take a leading role in operationalizing the big data analytics framework proposed in this study by its center, called the Rwanda Biomedical Center (RBC).

In future studies, the explainable machine learning models shown in Figure 7 will be developed, trained, and tested using large diabetes datasets from Rwanda, and patient empowerment through chatbot-based tools for improving diabetes education, self-management, follow-up, and monitoring, and for promoting a diabetes-friendly diet using locally available food with a low glycemic index. The full framework will be implemented in close collaboration with the Rwanda Biomedical Center, which will oversee how the models perform in practice, support clinician training, and help scale the proposed big data framework across healthcare systems in Rwanda.

## Acknowledgement

The authors express their gratitude to the University of Rwanda and Stockholm University for their support and facilitation during this study. This research was funded by the UR-Sweden Program for Research and Capacity Building Program through a scholarship to Silas Majyambere. We are grateful to our participants for their time and fruitful contributions during the workshop group discussion. Grammarly was used as a text formatting tool while writing the manuscript. All authors contributed equally to writing the manuscript. Silas was responsible for research design, data collection, methodology, and drafting the write-up. Tony and Celestin supervised the study, enhanced the manuscript, and analyzed the results. Workneh, leveraging his Big Data applications, assisted in designing the BDA Framework for Diabetes Management. All authors have reviewed and approved the final version of the manuscript.

## Ethical Statement

This research does not require ethical approval. The participants have agreed to waive the requirement for signing a consent form, as no identifiable personal information was collected or used in this study.